\documentstyle[11pt,newpasp, twoside, epsf]{article}

\begin{document}

\title{Dwarf Elliptical Galaxies in the Perseus Cluster}
\author{J.S. Gallagher (1), C.J. Conselice (1), R.F.G. Wyse (2)}
\affil{1. U. Wisconsin-Madison, 2. Johns. Hopkins U.}

\begin{abstract}

Dwarf ellipticals have the lowest stellar densities of any galaxies, but 
paradoxically are most common in the densest regions of the
universe, and are especially frequent in rich clusters of galaxies. Simple 
estimates suggest that low luminosity dwarf ellipticals should be subject to 
substantial tidal heating in rich clusters if most of their mass is in the 
form of normal stars. We have therefore undertaken a program to image 
dwarfs in clusters with the 3.5-m high performance WIYN and ARC and smaller, 
wide field telescopes. Our objectives include testing for evidence of 
tidal disruption in the form of asymmetries and searching for evidence of 
recent star formation which might be associated with the production of 
elliptical dwarfs from infalling field galaxies.

\end{abstract}


\section{Overview}

 Dwarf elliptical galaxies appear in abundance in galaxy clusters 
containing strong gravitational tides.  These
low luminosity, low surface brightness objects dominate clusters by number, 
giving them very steep luminosity functions.   A basic question to ask is: 
where did these dwarfs come from, and how can they survive in rich clusters?
Also, what is the relationship between the formation and evolution
of giant galaxies and dwarfs?  

  To help answer these questions, we obtained images of Perseus,
a rich galaxy cluster with a radial velocity of 5500 km/s 
(D=75 Mpc w/  H$_{0}$ = 75 km/s/Mpc).  Morphological identification of dEs is 
tricky, and most previous
studies pick their dE sample based on colors, typically 1.2$<$(B-R)$<$1.6.  
However, given the relative proximity of Perseus, it is possible to reverse 
this procedure - that is, to determine morphologically the dE sample, and 
then derive physically properties on this morphological selected sample. 

 Our major preliminary conclusions for this study are: 

$\bullet$ Dwarf candidates in Perseus appear to have a wide range of colors, 
making any simple, one time formation scenario unlikely. 

$\bullet$ We find the dEs, including the nucleated dEs (dE,N), 
are consistently the same color throughout the dwarf. This indicates the 
stellar populations are likely homogeneous in these dwarfs.

  A Perseus cluster field of 35 kpc x 35 kpc showing the high number of dwarfs 
from a WIYN R-band observations is presented in Figure 1.
We are able to identify 85 candidate dEs
in our Perseus cluster images.  This gives a number density of 1000 
dEs/Mpc$^2$.  
The basic properties of the dwarfs in Perseus are as follows:  Absolute Magnitude range : -14$<$ M$_{R}$ $<$-11, colors : 0.67$< $(B-R)$ <$2.6, and
central surface brightness: 22 $<$ $\mu_{R}$ $<$25.

The surface brightness, and magnitudes are standard, and 
agree with other observations of dwarf ellipticals, but 
the colors of dEs in Perseus cover a wide range.   There is a slight
correlation between the absolute magnitude of the brighter dEs in Perseus and 
the (B-R) color (Figure 2).  The line in Figure two is the relationship
found between (B-R) and magnitude for dEs in the Coma cluster by
Secker et al. (1997).

  Nucleated dEs are a well established morphological sub-class, and appear as 
bright dEs with
a concentrated center.   These nucleated cores are generally
believed to be super massive star clusters that are usually about 50 pc
in size.   The color of these nucleated cores have typically been
found to be indistinguishable from the remainder of the galaxy. 
We verify this for the Perseus dEs.

  What are the origin of these dEs?  Following Merritt (1984) galaxies in 
clusters are subject to long term tidal forces from the cluster core; 
typically $\approx$ 10$^{14}$ M$_{\odot}$ within a radius of 200-300~kpc. 
These should tidally truncate extreme dE dwarfs at $R \sim$3-10~kpc;
 we find the largest dEs in Perseus and Coma to be about this size.  
Tides therefore may control dE sizes near cluster cores.  
Likewise, galaxies on high eccentricity orbits 
cross a cluster core in a time comparable to internal orbital periods for 
stars in their outer envelopes.  These systems may experience tidal 
shocking, resulting in dynamical heating of the dwarf and possibly removal
of mass.
  Collisions from impact approximation 
models predict episodic heating and some mass loss.
Sometimes collisions will occur with low relative velocities; these can be
more damaging, and possibly produce dEs with a range of colors
from small disk galaxies.

\begin{figure}
\plottwo{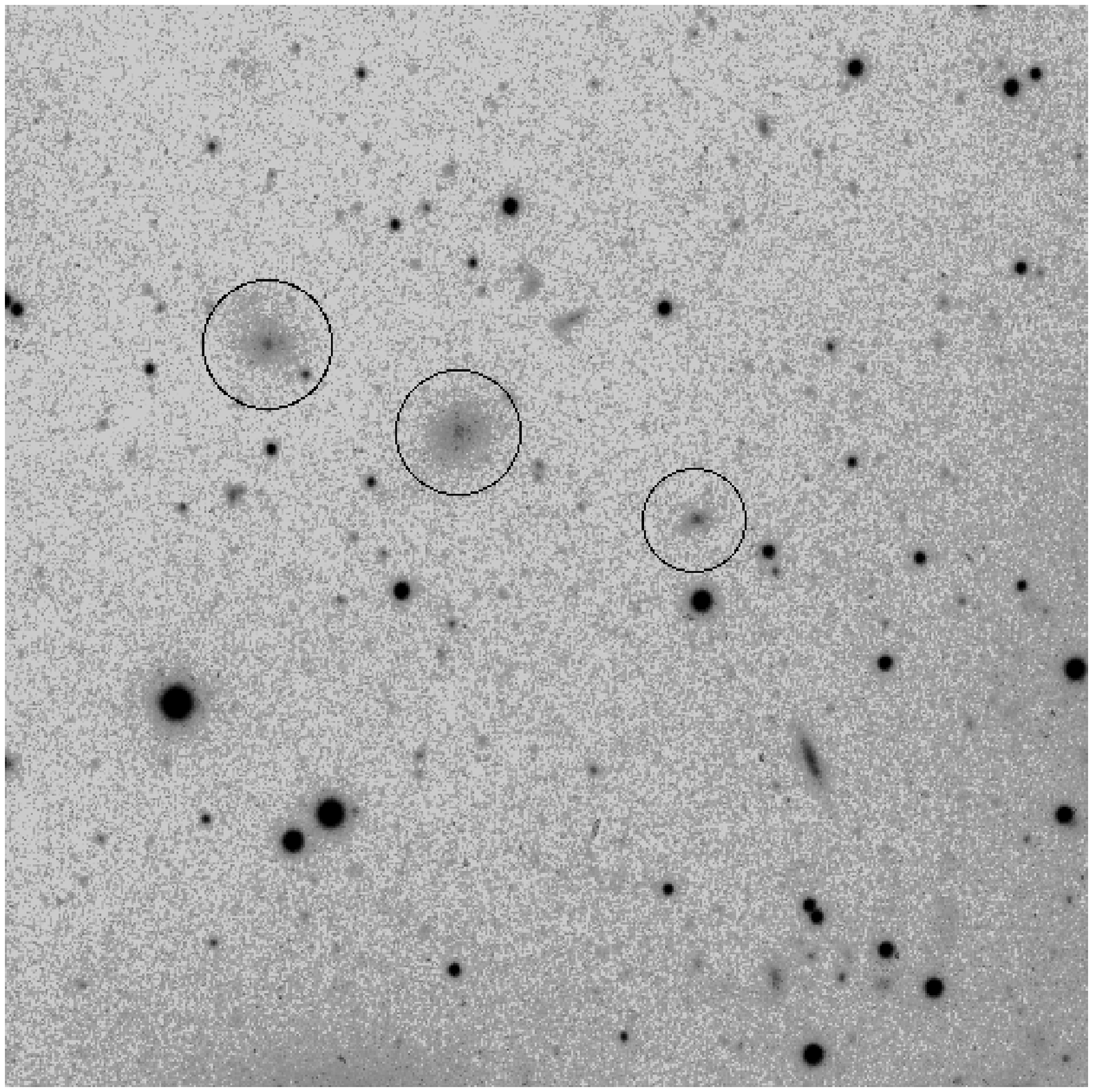}{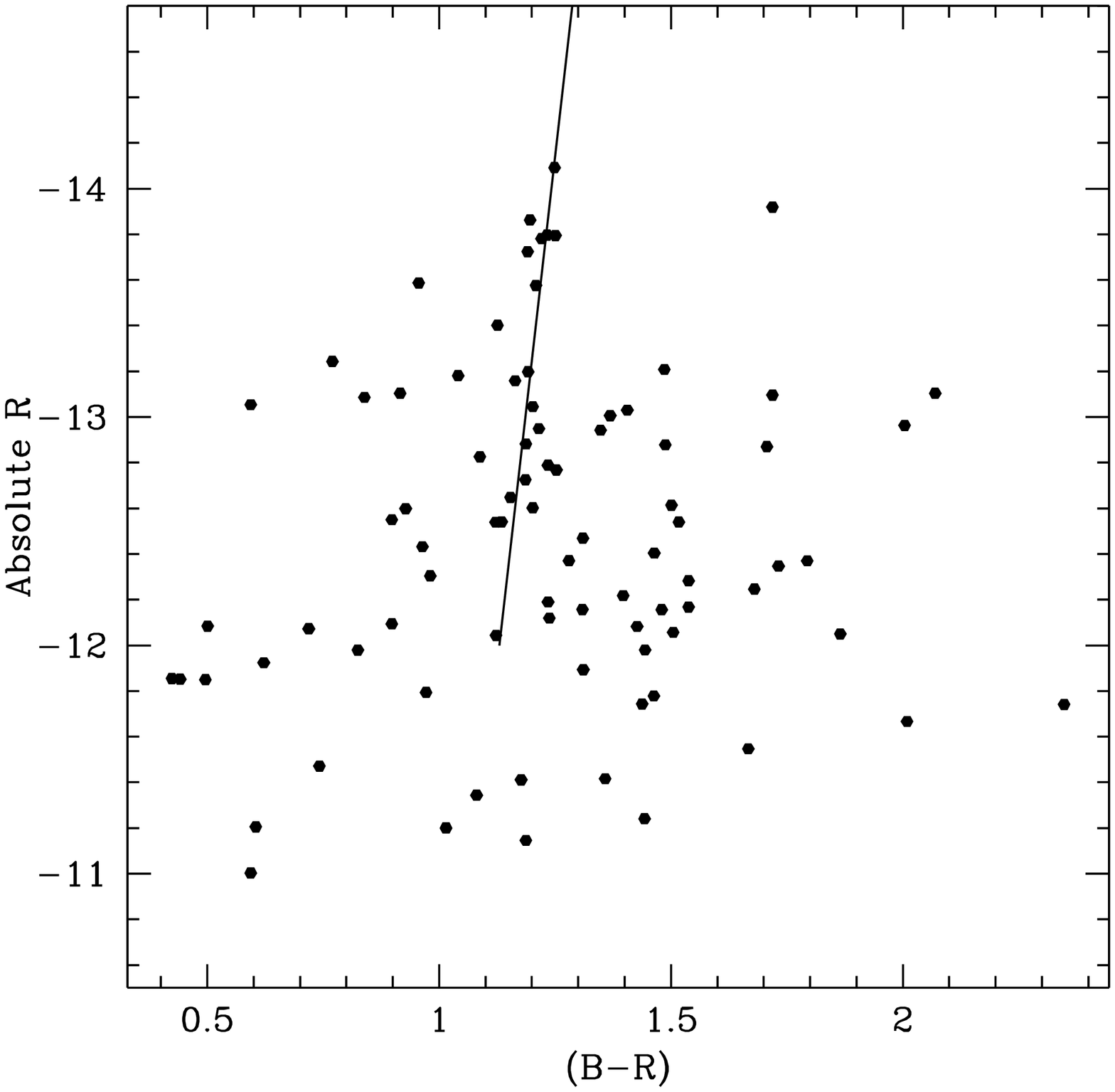}
\caption{\footnotesize{Examples of Perseus cluster dEs from a WIYN image, and our preliminary dE CMD.}}
\end{figure}


\begin{references}

\reference Secker, Harris, Plummer, 1997, PASP, 109, 1377
\reference Merritt, D., 1984, ApJ, 276, 26
\end{references}
\end{document}